\pdfoutput=1
\documentclass[11pt]{article}
\usepackage[top=2.5cm, bottom=2.5cm, left=2.5cm, right=2.5cm]{geometry}
\usepackage[english]{babel}
\usepackage{algorithm}
\usepackage[noend]{algpseudocode}
\usepackage{amssymb,amsmath,amsthm,amsfonts}

\usepackage{subcaption}

\usepackage{enumerate}
\usepackage{multicol}
\usepackage{algorithm}
\usepackage[noend]{algpseudocode}
\usepackage{array}
\usepackage{authblk}%multiple affiliations
\usepackage{multirow}
\usepackage{graphicx}
\usepackage{float}
\usepackage{color}
\usepackage{libertine}
\usepackage[authoryear]{natbib}

\usepackage{mathtools}

\usepackage{hyperref}

\begin{document}

\title{\textbf{Local migration quantification method for scratch assays}}

\author[1,2]{\small Ana Victoria Ponce Bobadilla\footnote{anavictoria.ponce@iwr.uni-heidelberg.de}}
\author[3]{\small Jazmine Ar\'evalo}
\author[3]{\small Eduard Sarr\'o}
\author[4]{\small Helen Byrne}
\author[4]{\small Philip K Maini}
\author[1,2]{\small Thomas Carraro}
\author[5,6]{\small Simone Balocco}
\author[3,7,8]{\small Anna Meseguer}
\author[9,10,11,12]{\small Tom\'as Alarc\'on \footnote{Talarcon@crm.cat}}

\affil[1]{ Institute for Applied Mathematics, Heidelberg University, 69120 Heidelberg, Germany}
\affil[2] {Interdisciplinary Center for Scientific Computing (IWR), Heidelberg University, 69120 Heidelberg, Germany}
\affil[3]{Renal Physiopathology Group, CIBBIM-Nanomedicine, Vall d'Hebron Research Institute, Barcelona, Spain}
\affil[4]{Wolfson Centre for Mathematical Biology, Mathematical Institute, University of Oxford, Oxford OX2 6GG, UK}
\affil[5]{Dept. Matematics and Informatics, University of Barcelona, Gran Via 585, 08007 Barcelona, Spain}
\affil[6]{Computer Vision Center, 08193 Bellaterra, Spain}
\affil[7]{ Departament de Bioqu\'imica i Biologia Molecular, Unitat de Bioqu\'imica de Medicina, Universitat Aut\`onoma de Barcelona, Bellaterra; Spain}
\affil[8]{ Red de Investigaci\'on Renal (REDINREN), Instituto Carlos III-FEDER, Madrid, Spain}
\affil[9]{ ICREA, Pg. Llu\'is Companys 23, 08010 Barcelona, Spain}
\affil[10]{ Centre de Recerca Matem\`atica, Edifici C, Campus de Bellaterra, 08193 Bellaterra (Barcelona), Spain}
\affil[11]{Departament de Matem\`atiques, Universitat Aut\`onoma de Barcelona, 08193 Bellaterra (Barcelona), Spain}
\affil[12]{ Barcelona Graduate School of Mathematics (BGSMath), Barcelona, Spain}
\date{}
\maketitle
\vspace{-25pt}

\begin{abstract}
\textbf{Motivation:} The scratch assay is a standard experimental protocol used to characterize cell migration. It can be used to identify genes that regulate migration and evaluate the efficacy of potential drugs that inhibit cancer invasion. In these experiments, a scratch is made on a cell monolayer and recolonisation of the scratched region is imaged to quantify cell migration rates. A drawback of this methodology is the lack of its reproducibility resulting in irregular cell-free areas with crooked leading edges. Existing quantification methods deal poorly with such resulting irregularities present in the data.  \\
\textbf{Results:} We introduce a new quantification method that can analyse low quality experimental data. By considering in-silico and in-vitro data, we show that the method provides a more accurate statistical classification of the migration rates than two established quantification methods. The application of this method will enable the quantification of migration rates of scratch assay data previously unsuitable for analysis. \\
\textbf{Availability and Implementation:}
The source code and the implementation of the algorithm  as a GUI along with an example dataset and user instructions, are available in \url{https://bitbucket.org/anavictoria-ponce/local_migration_quantification_scratch_assays/src/master/}. The datasets are available in \url{https://ganymed.math.uni-heidelberg.de/~victoria/publications.shtml}.
\end{abstract}

\section{Introduction}

Cell migration plays a fundamental role in developing and maintaining the organization of multicellular organisms, while aberrant cell migration is found in many pathological disorders like cancer and atherosclerosis \citep{friedl2009collective,li2013collective,roussos2011chemotaxis}. Scratch or wound healing assays are standard in-vitro methods for studying cell migration \citep{kramer2013vitro,liang2007vitro}. A scratch assay involves: growing a cell monolayer to confluence in a multiwell assay plate; creating a ``wound'', a cell-free zone in the monolayer, into which cells can migrate; monitoring the recolonisation of the scratched region to quantify the cell motility \citep{liang2007vitro}. This experimental technique is routinely used to identify genes that regulate cell migration \citep{simpson2008identification} and to evaluate the efficacy of potential drugs that inhibit cancer invasion \citep{hulkower2011cell} by comparing the migration rates of replicates of scratch assays under the same cell conditions but different drug levels. Given its key role in assessing compounds for clinical use, it is important to develop robust quantification methods that accurately compare migration rates of different scratch assays.

Although the scratch assay is a standard procedure, there is no standard method for quantifying cell migration \citep{topman2012standardized}. The most common methods focus on the wound size evolution  \citep{grada2017research, masuzzo2016taking}. First, the leading edges of the spreading cell population are detected by manual tracking or by automated image analysis software, such as Image J \citep{ferreira2012imagej}, Tscratch \citep{gebaeck2009tscratch} or Matlab's Image Processing Toolbox. Cell migration is then quantified by: the percentage difference in the wound sizes at different time points \citep{ranzato2011wound,walter2010mesenchymal},  the wound size at specific time points \citep{buth2007cathepsin}, or the slope of a linear approximation to the evolution of the wound area \citep{jonkman2014introduction}.

\begin{figure}
\centering
  \includegraphics[width=.8\linewidth]{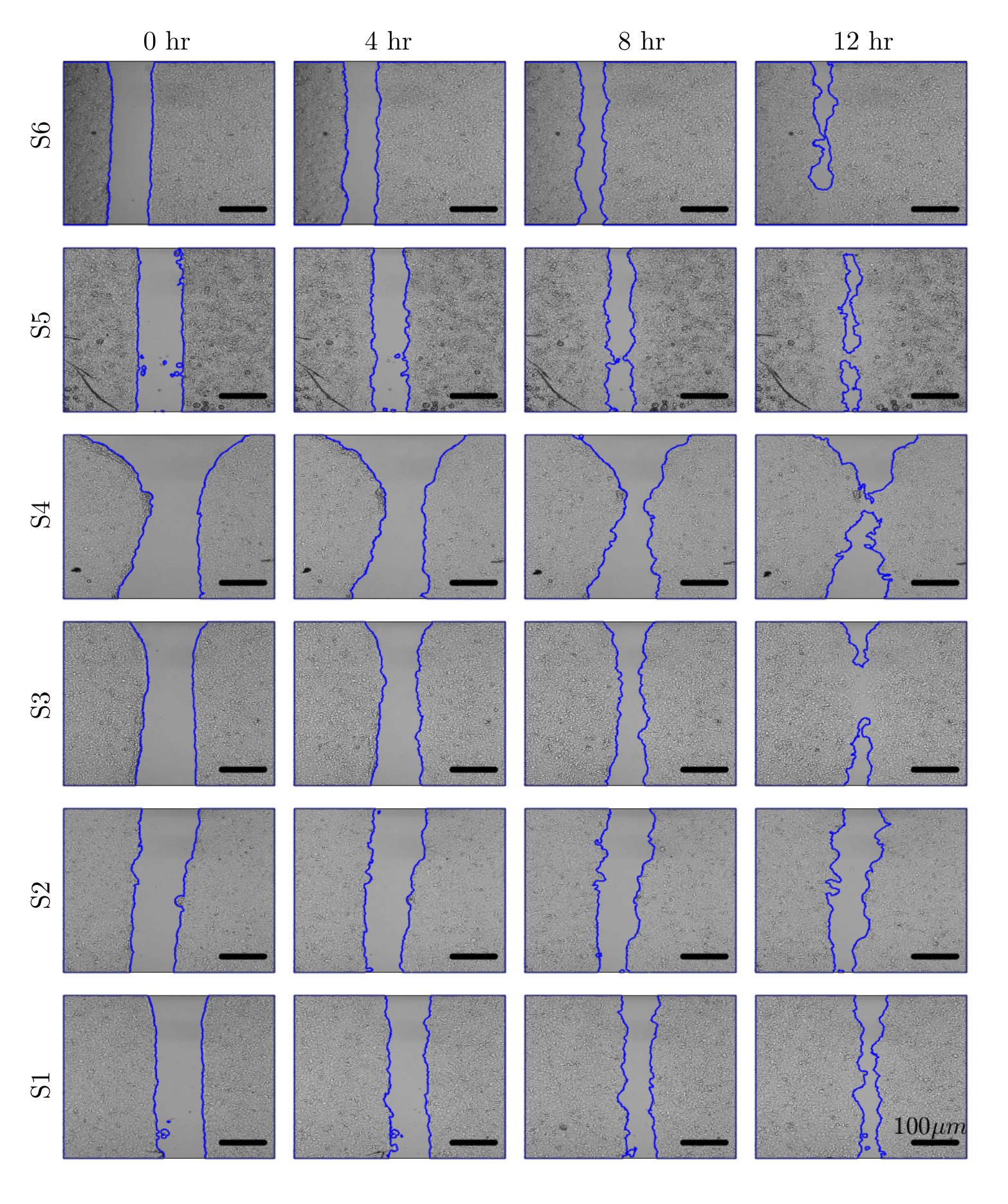}
  \caption{ The evolution of a representative scratch from each of the six cell groups (S1-S6) from our experimental data is plotted at time 0 hr, 4 hr, 8 hr and 12 hr.  In each image, the leading edges have been detected by applying the segmentation algorithm. The detected interfaces/cell fronts are plotted in blue.
 \label{fig:segmentation_example}}
\end{figure}

Current quantification methods do not perform well when the two borders of the scratch are not straight. A major drawback of the scratch assay methodology is the lack of reproducible wounding that results in non-uniform cell-free areas with irregular leading edges, as can be seen in Figure~\ref{fig:segmentation_example}. Migration rate measurements have been shown to be sensitive to the initial degree of confluence \citep{jin2016reproducibility}, the experimental design \citep{jonkman2014introduction}, and the initial geometry of the wound \citep{jin2017role}.

While considerable effort has focused on improving and standardizing the automated detection of the leading edge  \citep{gebaeck2009tscratch,zaritsky2011cell,zaritsky2013benchmark}, few authors have focused on standardizing the migration quantification method. Since migration rate measurements have been shown to be sensitive to the choice of the detection method \citep{treloar2013sensitivity,zaritsky2011cell}, it is important to consider the same edge detection method when comparing the performance of quantification methods. While some authors have compared different migration quantification methods \citep{vargas2015robust}, the influence of the detection algorithm choice has not yet been considered.

In this work, we introduce a new method that quantifies the velocities in the perpendicular direction of the cell fronts. Irregular leading edges are accounted for by approximating the front by a piecewise constant function, which is constant over windows with a fixed size $w^*$. We assume that in each window, the contour moves with constant speed in the perpendicular direction until the two leading edges meet. We approximate this velocity by the gradient of a linear approximation to the front position evolution in this window. The window size, $w^*$, is chosen to be the one that best fits a constant velocity profile. The migration in the scratch assay is then characterized by the slopes of a series of linear approximations to the interface evolution in these windows.

In order to validate the quantification method and compare it to other methods in a controlled situation, we use an agent-based model to produce in-silico scratch assay data. In the agent-based model, the agents move and proliferate with defined probabilities. We investigate how the velocity distribution, determined by our quantification method, is affected by cell motility and proliferation. In this controlled situation, we compare our quantification method with two other methods: the \textit{percentage wound area method} which is widely used and the \textit{closure rate method}, introduced by \cite{jonkman2014introduction}. By using classification tests in which we simulate an experimental setting based on in-silico data, we show that our method outperforms these methods.

We then apply the quantification methods to in-vitro experimental data. We perform statistical tests to analyse differences between one group with respect to all the others. Our quantification method is able to detect significant differences with respect to some of the other groups while the closure rate method was able to detect only one difference and the area method did not detect any difference.

The paper is organized as follows: in Section \ref{subsec:cellculture} we describe our experimental system and in Section \ref{subsec:agentbased} we present the agent-based model that we use to simulate the in-vitro process. In Section \ref{subsec:migration_quantification} we introduce a new migration quantification method for scratch assays and describe the two quantification methods with which we compare it. By considering the agent-based model, in Section \ref{subsec:senstivity_analysis_abm}, we investigate how the quantification method is affected by cell motility and proliferation. In Section \ref{subsec:velocity_performance} we show that the method correctly classifies cells with different motility and proliferation parameters. In Section \ref{subsec:comparison_quant_methods} we show that our method outperforms the other methods. Finally, in Section \ref{subsec:classification_experimental_data}, we apply the method to an experimental data set and find statistically significant differences in the migration rates when we compare samples from different groups of cells; the percentage wound area and closure rate methods were unable to detect these differences.

\section{System and methods}

\subsection{Cell culture and wound healing assay }
\label{subsec:cellculture}

Six site-specific mutations in a latent transcription factor that regulates downstream genes involved in essential
biological processes, including migration, were generated. Mutants S1, S2, S3, S4, S5 and S6 were then transduced into a human renal carcinoma cell line, 769-P (ATCC CRL-1933), through lentiviral particles. The 769-P mutants were cultured in Dulbecco's Modified Eagle's Medium (DMEM) (\#42430, Gibco) supplemented with 10\% of fetal bovine serum (FBS) (\#10270, Gibco), 1\% of sodium pyruvate solution 100 mM (\#03-042-1B, Biological Industries) and 1\% of antibiotic-antimycotic solution 100X (\#15240, Gibco). Cells were maintained at 37$^\circ$C in 5\% CO$_2$.

For the wound healing assay, the 769-P mutants (S1-S6) were seeded at $0.025 \times  10^6$ cells per well in a 2 well silicone insert with a defined cell-free gap (Ibidi \#81176, Germany), incubated and allowed to grow for $48$ hr. Once the cells reached 100\% confluence, the culture insert was removed and the area that remained clear of cells was quantified for $24$ hr using the Live Cell-R Station (Olympus). Digital images were obtained every 30 minutes.

Data consisted of 24 wound healing assays: four replicates for each of the six groups (S1-S6). Each assay consisted of 48 images. The imaged region size was $500\times 500\ [\mu m]^2$.

\subsection{Agent-based model of the scratch assay}
\label{subsec:agentbased}

We consider an agent-based model that has been previously used to simulate in-vitro cell cultures  \citep{johnston2016quantifying,johnston2014much,simpson2010cell}. The simulation domain is a two-dimensional square lattice, with the same dimensions as the experimental images: $[0,D]\times [0,D]$ where $D=500\ \mu m$. The lattice spacing, $\Delta$, which is interpreted as the average cell diameter, is set to $10\ \mu m $ unless otherwise specified.

In this model each agent can either proliferate or move within the simulation domain. We consider an end time of $T=24$ hr and an update time of $\tau=0.1$ hr.
 We include crowding effects by assuming that each lattice site is occupied by at most one cell. A cell with centre at $(x,y)$ is said to be at $(x,y)$. Zero flux boundary conditions are imposed. The simulation algorithm is summarized below:

\begin{figure*}[!tpb]
  \centering
  \includegraphics[width=\linewidth]{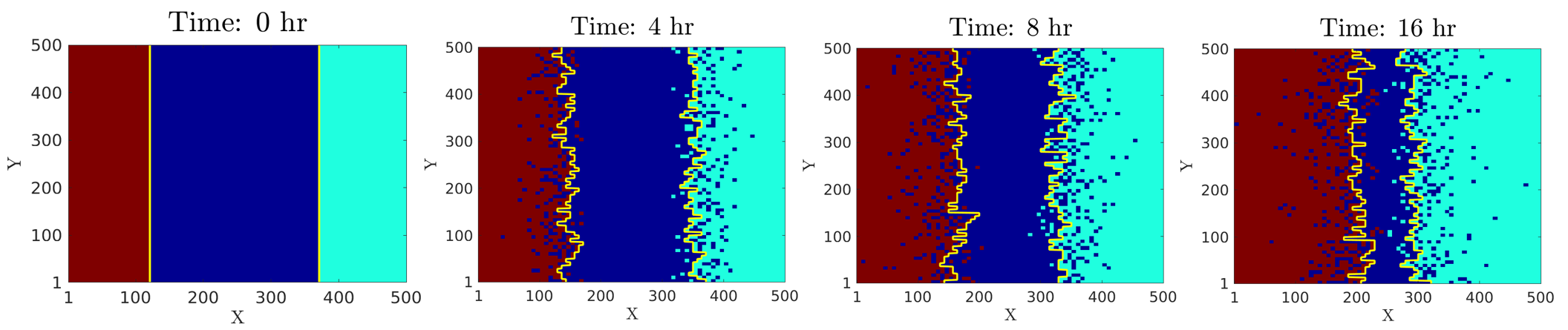}
  \caption{Evolution of an agent-based simulation. We considered an idealised initial condition and fixed the motility and proliferation parameters so that $p_m=0.3$ and $p_p=0.01$, respectively. The recolonisation of the wounded region is shown at times  $t=0,4,8$ and $16$ hr. The two cell monolayers are plotted with different colours (red and turquoise), while the area devoid of cells is coloured blue. The leading edges detected by the segmentation algorithm are plotted in yellow.   \label{fig:scratch_evolution}}
\end{figure*}

\begin{enumerate}[(a)]
\item \textit{Initialization.} We consider an idealised initial condition unless otherwise specified: lattice points $(x,y)$, for which $x<D/ 4$ or $x>3D/ 4$, are occupied by an agent (See Figure~\ref{fig:scratch_evolution}).
\item \textit{Update algorithm. } Let $N(t)$ denote the number of agents at time $t$. To update the agent-based model at time $t$ to the next simulation time $t+\tau$, we do the following:
 \begin{enumerate}[1. ]
 \item First, $N
 (t)$ agents are chosen sequentially at random and given the opportunity to move. An agent at $(x,y)$ attempts to move with probability $p_m$ to $(x\pm \Delta, y)$ or $(x, y\pm\Delta)$, with the target site chosen with equal probability.
 \item $N(t)$ agents are then selected sequentially at random again and given the opportunity to proliferate. An agent at $(x,y)$ attempts to proliferate with probability $p_p$ and places its daughter agent at  $(x\pm \Delta, y)$ or $(x, y\pm\Delta)$, target sites being chosen with equal probability.
 \end{enumerate}
At each update time, agents can move and/or proliferate only if the target site is vacant. Since typical estimates of the cell doubling time are approximately 15-30 h \citep{maini2004traveling, simpson2013quantifying}, whereas the time required for a cell to move a distance equal to its diameter is of 10 min \citep{khain2011collective}, we vary the motility and proliferation probabilities in the ranges $p_m\in[0,1]$ and $p_p\in[0,0.01]$, respectively. The agent-based model parameters are presented in Table \ref{table:parameters_abm}.
\end{enumerate}

In Figure~\ref{fig:scratch_evolution} we plot the evolution of a typical realisation of the agent-based model simulation for which the motility and proliferation parameters are given by $p_m=0.3$ and $p_p=0.01$, respectively.

\begin{table}[h!t]
\center
 \begin{tabular}{| c| c | c| c|  }\hline Parameter & Description & Value & Units  \\ \hline
 $p_m$ &  Motility probability & $0.3$ & N/A \\ \hline
 $p_p$ & Proliferation probability & $0.1$ & N/A \\ \hline
 $T$ & End time & 24 & hr \\ \hline
 $\tau$ & Time step & $0.01$ &  hr \\ \hline
 $D$ & Domain length & $500 $ & $\mu m$ \\ \hline
 $\Delta$ & Lattice spacing & $15$ &$ \mu m$ \\ \hline
\end{tabular} \caption{Parameter values of the agent-based model\label{table:parameters_abm}}
\end{table}

\subsection{Automatic contour segmentation}

The leading edges of the cell monolayers from the experimental images are detected by applying a segmentation algorithm based on the Growcut method \citep{vezhnevets2005growcut}. The method is a robust technique, already employed in several computer vision applications, that performs a binary image segmentation.

The Growcut algorithm requires the initial specification of a subset of pixels from each type of region: cell monolayer and unoccupied space; these pixels are referred to as \textit{seeds}. The seeds should be located far from the leading edges, where all the pixels of such an area belong to one of the two classes. The algorithm evolves as follows: at each iteration, the pixels surrounding the initial seeds are assigned to one class or the other, adjusting the size of each region. The classification depends on the similarity of the pixel intensity with respect to the pixel intensity of the seeds. When the segmentation process is finished, the interface between the two regions represents the segmented cell fronts.

In our implementation, the seeds are chosen as follows: for the cell region, the Canny and Roberts edge contour methods \citep{canny1987computational, shrivakshan2012comparison} are used to select the pixels with the highest variability, corresponding to the cell contours. For the background region, the seeds are set in areas having a low variability, defined as areas in which the pixel intensity has a standard deviation less than 500.

After applying the detection algorithm to each image, we have a record of the positions of the left and the right interfaces at each time where the image was taken. At each vertical position, the interface is consider to be the closest pixel to the wound.

\subsection{Migration quantification methods }\label{subsec:migration_quantification}

We first introduce the two established quantification methods for scratch assays. Then, we introduce a new method that quantifies the x-component of the velocity of the leading edge of the cell monolayer.

\subsubsection{Percentage wound area method }

The most common quantification method, which we refer to as the \textit{area method}, assesses the migration in an indirect manner. In the course of the experiment, the wound area percentage, $\hat{A}(t)$, is tracked:
\begin{equation*}
\hat{A}(t):=\frac{A(t)}{A(0)}\times 100 \%
\end{equation*}
where $A(t)$ is the wound area at time $t$ and $A(0)$ is the initial area. The migration rate is then indirectly evaluated as the percentage wound area at a specific time point. Typically when the migration rates of a cell line under different experimental conditions are compared, the percentage wound areas of the samples at specific time points are compared. The time points of comparison are not standardized and vary across studies \citep{walter2010mesenchymal,ruiz2017amniotic,gorshkova2008protein}.
The choice of time points can affect the comparison, making the results uncertain.

\subsubsection{Closure rate method}

\cite{jonkman2014introduction} proposed a method for quantifying cell migration with respect to the slope of a linear approximation to the evolution of the wound area. We refer to this method as the \textit{closure rate method}. The evolution of the wound area $A(t)$ is first approximated by a linear function:
\begin{equation}
A(t)\approx m\times t +b\label{eq:area_evolution}
\end{equation}
where $m$ and $b$ are real scalars. The wound area is assumed to be the length of the field-of-view ($l$) times the width of the gap ($W(t)$). Since $l$ is constant during the course of the experiment, Equation \eqref{eq:area_evolution} becomes:
\begin{equation}
\frac{dA}{dt}\approx l\times \frac{dW}{dt}.\label{eq:areaevol}
\end{equation}
The migration rate, $C_r$, is defined to be half of the width closure rate
\begin{equation}
C_r:= \frac{1}{2} \frac{dW}{dt}.\label{eq:cr_definition_from_width}
\end{equation}
From Equations \eqref{eq:areaevol} and \eqref{eq:cr_definition_from_width}, we have
\begin{equation}
C_r=\frac{\vert m\vert }{2\times l }.
\end{equation}

\subsection{Proposed quantification method: velocity method }

\begin{figure*}[!tpb]
  \centering
  \includegraphics[width=\linewidth]{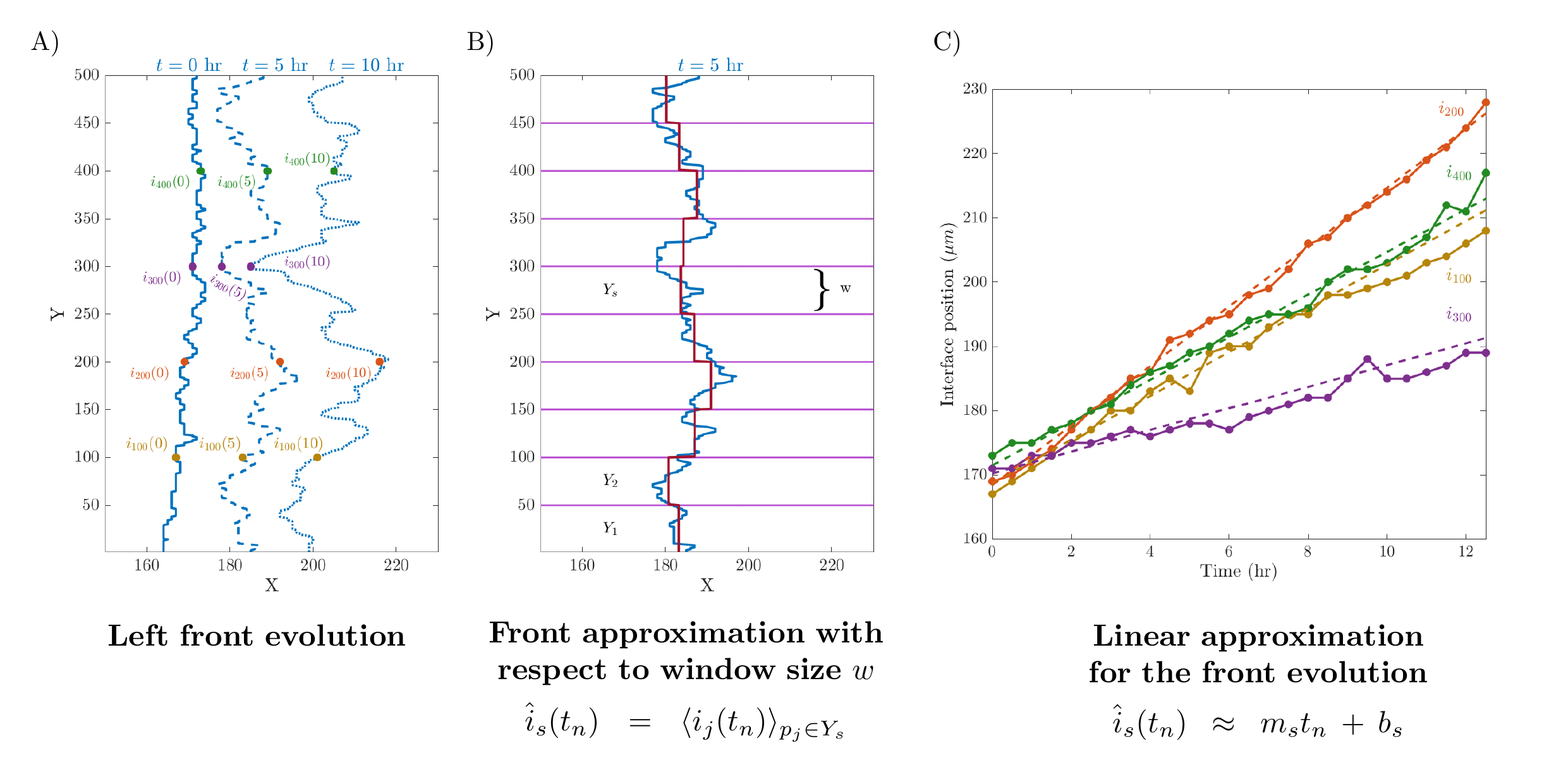}
  \caption{Linear approximation of the front evolution with respect to window size $w$. A) To introduce the notation, the left front position at times $t=0, 5,$ and $10 $ hr are plotted in blue. The solid line corresponds to $t=0$ hr, the dashed line to $5$ hr and the dotted line to $10$ hr. The front position at the 100, 200, 300 and 400 y-coordinates for these times is plotted in different colours: yellow, orange, purple and green, respectively. B) The left front at $t=5$ hr is approximated by a window size $w$. $Y$ is partitioned into $Y_s$ segments each with length $w$. A magenta horizontal line delimits each segment. The front position is plotted in blue and the approximated front position, taken as an average over each $Y_s$, is plotted in red. C) The interface evolution at the 100, 200 300 and 400 y-coordinates and the linear approximation with respect to the window size $w=12$ are plotted in full lines and dashed lines, respectively. The window size $w=12$ is the window size that maximizes the objective function \eqref{eq:global_fit}.  \label{fig:linear_approximation}}
\end{figure*}

We propose a new strategy for quantifying front migration in a scratch assay by a set of representative velocities. We denote by $t_0,\ldots, t_N$, the times at which data are collected.  Let $X\times Y$ represent the square domain of the processed image, $X=Y=\{1,\ldots, D\}$ where D is the number of pixels. For each $j\in Y=\{1,\ldots, D\}$, we denote the interface position in the horizontal direction, at the $j-$th vertical position and at time point $t_n$, as $i_j(t_n)$ where $1\leq i_j \leq D$. (see Figure~\ref{fig:linear_approximation} A) for a schematic representation ).

To determine the velocities, we perform a linear approximation to the front evolution with respect to a window size $w$. The linear approximation is defined in two steps:
\begin{enumerate}
\item \hspace{.05cm} First, the front position is approximated with respect to the window size, $w$. $Y$ is divided into $M=D/w$ $Y_s$ segments of length $w$. The front position in each segment $Y_s$ is approximated by its mean position,
\begin{equation}
\hat{i}_s(t_n)=\langle i_j(t_n) \rangle_{i_j\in Y_s}.
\end{equation}
This procedure is illustrated in Figure~\ref{fig:linear_approximation} B).
\item \hspace{.05cm}The front evolution along each window is approximated by a linear regression
\begin{equation}
\hat{i}_s(t_n)\approx m_s t_n+b_s. \label{eq:interface_approximation}
\end{equation}
In Figure~\ref{fig:linear_approximation} C) the solid lines represent the front evolution at selected y-coordinates; the dashed lines represent the corresponding linear approximations for a window of size $w=16$.
\end{enumerate}

Performing this approximation for the left and right interfaces, we obtain a set of velocities $ \{\vert m_s\vert \}_{s=1}^{2M}$ which we refer to as the windowed velocities for window size $w$.

Given a window size $w$, we expect the left and right windowed velocity distributions to be similar, since both fronts were initially part of the same cell monolayer. However, for window sizes smaller than the average cell size, the distributions are significantly different. This is because the scale on which the velocities are approximated is then much smaller than the cell size scale: the individual velocity of each cell at the front is counted multiple times and its value is over represented, producing a bias in the overall windowed velocity distribution. In practice, we choose a window size for which the left and right windowed velocity distributions are similar.

The key step of our method is to determine the optimal window size to perform the linear approximation \eqref{eq:interface_approximation}. We use two criteria to select the optimal window size, $w^*$: (i) fitness of the approximation, and (ii) similarity of the left and right windowed velocity distributions. We consider an objective function, $F(w)$, that allows us to find the optimal window size with respect to these two criteria. The objective function has three terms:
\begin{equation}
F(w)=Fit_{resid}(w)+Fit_{Rsquared}(w)+Fit_{KS_{distance}}(w). \label{eq:global_fit}
\end{equation}
\begin{itemize}
\item $Fit_{resid}(w)$ measures the discrepancy between the interface evolution and the linear approximation (Equation \eqref{eq:interface_approximation}).
\item $Fit_{Rsquared}(w)$ considers the coefficient of determination, $R^2$, which describes how well the evolution variance is explained by the linear approximation \citep{heiberger2015statistical}.
\item $Fit_{KS_{distance}}(w)$ is a distance function derived from the Kolmogorov-Smirnov statistic for the two-sample Kolmogorov-Smirnov (K-S) test \citep{heiberger2015statistical} that calculates the distance between the left and right front windowed velocity distributions.
\end{itemize}
The terms are scaled such that the window size that maximises the objective function gives the best fit and has the left and right velocity distributions which are closest to each other. Detailed information about the objective function can be found in Supporting Information 1.

The steps used to determine the set of representative velocities are summarized in Algorithm \ref{alg:quantmethod}.

\begin{algorithm}[H]
\caption{: Velocity quantification method }\label{alg:quantmethod}
\begin{algorithmic}[1]
\State Determination of the optimal window size for the linear approximation using the objective function \eqref{eq:global_fit}
\begin{equation}
w^*=\underset{1 \leq w \leq D}{\max}F(w)
\end{equation}
where $F(w)$ is given by equation \eqref{eq:global_fit}.
\State Linear approximation with respect to the window size $w^*$ for the evolution of the left and right interfaces,
\begin{equation*}
\hat{i}_s(t_n)\approx m_s t_n+b_s
\end{equation*}
where $\hat{i}_s(t_n)=\langle i_j(t_n) \rangle_{i_j\in Y_s}$, $ Y=\bigcup\limits_{s=1}^M  Y_s $ in which $\vert Y_{s}\vert=w^*$ and $M=D/w^*$.
\Ensure  $ \{\vert m_s\vert \}_{s=1}^{2M}$ is the representative set of velocities that quantify the migration in the scratch assay.
\end{algorithmic}
\end{algorithm}

\subsection{Classification test}

In order to assess the performance of the three quantification methods in a controlled way, we use the agent-based model to generate in-silico scratch assays. In particular, we compare the ability of the different methods to distinguish between cell populations with different proliferation and motility parameters. We consider the following classification test:
\begin{enumerate}
\item \hspace{.05cm}We fix a focal parameter combination $\hat{P}=(p_{\hat{m}},p_{\hat{p}})\in [0, 1] \times [0, 0.01]$ and run $n$ simulations of the agent-based model using these parameter values.
\item \hspace{.05cm}We decompose the parameter space of motility and proliferation probabilities $[0, 1] \times [0, 0.01]$ into a regular $11\times 11$ grid with 121 parameter pairs $(p_m,p_p)$. For each parameter combination, we run $n$ simulations of the agent-based model.
\item \hspace{.05cm}We calculate the migration rate of all simulations using the three quantification methods. The migration measurements are windowed velocities, closure rates or areas at specific time points, depending on the quantification method.
\item \hspace{.05cm}For each quantification method, we determine whether the migration measurements of each sampled parameter combination $(p_m,p_p)$ are statistically significantly different from the ones from those for the focal parameter pair $\hat{P}$. We perform two tests: the two-sample Kolmogorov-Smirnov test and the unpaired two-sample t-test, which we refer as the K-S test and t-test, respectively. We fix a $p$-value $ < 0.05 $ to define statistical significance.
\end{enumerate}
We consider a K-S test and a t-test to test for differences at the distribution level and in the mean. We test our data for normality and in case the migration measurements are not normally distributed, we consider a Wilcoxon rank-sum test. We account for stochasticity of the agent-based model by repeating this test for 20 times and analyse the mean and variance of the classification results.

 When applying the classification test to the velocity method, we consider a global optimal window size for determining the windowed velocities of the  simulations. In this way, we obtain the same number of windowed velocities for each simulation. To determine this global optimal window, we consider a weighted sum of the individual objective functions of each simulation (Supporting Information 1). When applying the classification test to the area method, we must specify the time point at which the wound areas are measured and compared. We fix the comparison time to be half the time it takes the leading edges to touch each other in the first simulation.

\section{Implementation}

The segmentation algorithm and the data analysis are implemented in MATLAB Version: 9.3.0.713579 (R2017b). The segmentation pipeline uses functions from Matlab's Image Processing Toolbox, the Grow Cut algorithm implementation found in
\url{http://freesourcecode.net/matlabprojects/56832/growcut-image-segmentation-in-matlab} and the normality tests implemented by \cite{oner2017jmasm}. The agent-based model is implemented in NetLogo \citep{tisue2004netlogo}. The output of the agent-based model simulation is a list of positions of the cells present at each update time. A C++ program is implemented to transform this output into a series of occupancy matrices, each saved in a separate file for each update time.  We do not apply the segmentation algorithm to the in silico images so the detection method does not affect the migration rate measurements. The occupancy matrix format provides the actual pixel classification of the in silico images into cell monolayer and empty space. The in silico images are then treated in the same way as the in vitro images. The source code of the programs are available in \url{https://bitbucket.org/anavictoria-ponce/local_migration_quantification_scratch_assays/src/master/}.

\section{Results}

\subsection{Exploration and validation of quantification method via in-silico data}

We first use the agent-based model to investigate how our quantification method is affected by cell motility and proliferation. Then, by applying the classification test, we investigate how well the method classifies cell populations with different motility and proliferation parameters in comparison with the other quantification methods.

\subsubsection{Sensitivity analysis}\label{subsec:senstivity_analysis_abm}

\begin{figure*}[!htb]
  \centering
     \includegraphics[width=\linewidth]{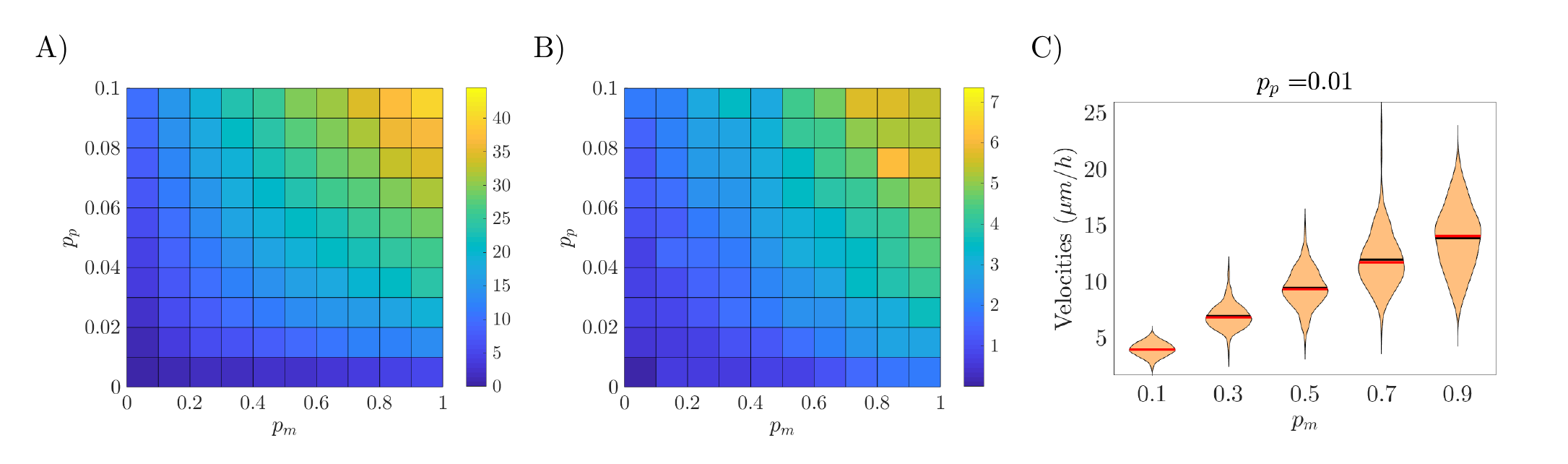}
 \caption{Sensitivity analysis of the agent-based model. We analyze the variability of the windowed velocities with respect to the proliferation and motility probabilities $(p_m,p_p)\in [0,1]\times[0,0.1]$. In each interval, 11  equally spaced points are considered that correspond to 121 $(p_m,p_p)$ parameter pairs. In A) and B), we plot the mean and the standard deviation of windowed velocities of 150 simulations under each of these 121 parameter pairs. In C), violin plots of the windowed velocity distributions while varying $p_m $ from 0.1 to 0.9 with a step of 0.2 and fixing the proliferation probability $p_p=0.01$, illustrate the general trend with the mean and standard deviation increasing as $p_m$ increases. \label{fig:sensitivity}}
  \end{figure*}

We investigate how the windowed velocities are affected by the rates of cell migration and proliferation. We vary the motility and proliferation probabilities for fixed initial conditions. We decompose the parameter space of motility and proliferation probabilities $[0, 1] \times [0, 0.01]$ into a regular $11\times 11$ grid with 121 parameter pairs $(p_m,p_p)$. For each parameter combination, 150 simulations were performed and the windowed velocities were calculated. The optimal window was calculated with respect to all simulations for the same parameter combination. In Figure~\ref{fig:sensitivity} A) we present a contour plot of the mean windowed velocity which shows how, as the probabilities increase, the mean velocity increases. A similar trend is observed for the standard deviation (see Figure~\ref{fig:sensitivity} B). The variation in the velocity distribution, as $p_m $ changes for increasing values of $p_m$ and a fixed proliferation probability $p_p=0.01$, can be seen in Figure~\ref{fig:sensitivity} C), where violin plots of the windowed velocities are shown.

\subsubsection{Classification performance}
\label{subsec:velocity_performance}

We suppose that the focal parameter combination, $\hat{P}=(\hat{p}_m,\hat{p}_p)$, takes values in $\{0.1,0.5,0.9\}\times\{0.01,0.05,0.09\}$ in order to test the classification for small, medium and high values of cell motility and proliferation in our parameter space. We consider $n=4$ simulations as the sample size for our test, so as to coincide with  experimental settings in which four samples are typically used. We repeat the classification test 20 times to produce results that account for the stochasticity of the system.

\begin{figure*}[!htb]
  \centering
   \includegraphics[width=\linewidth]{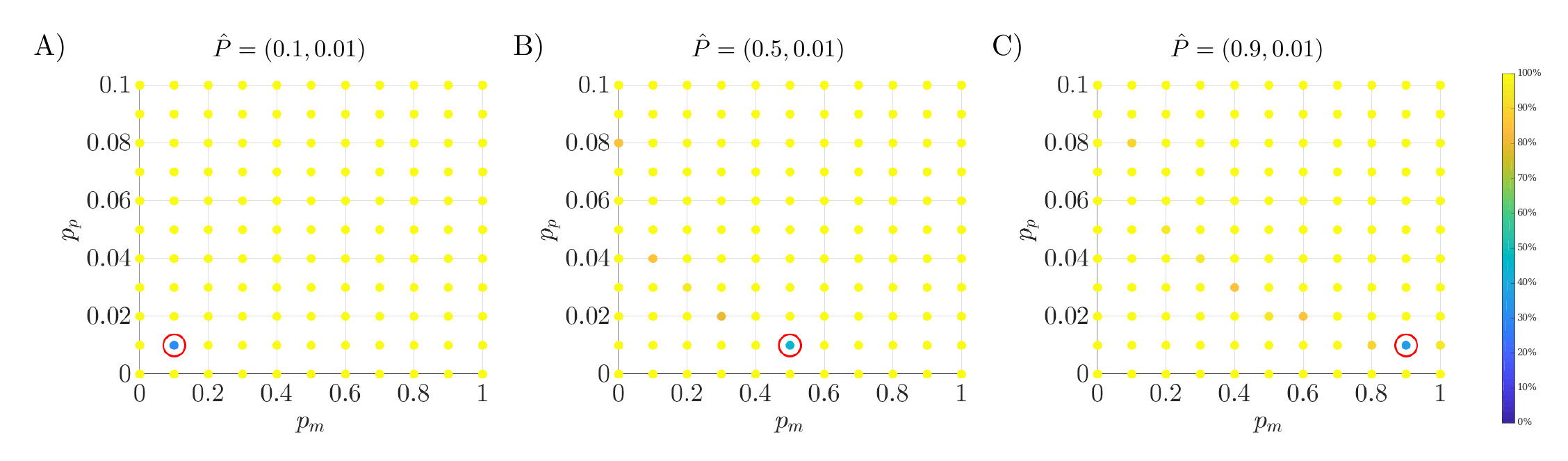}
 \caption{Plots of the mean behaviour of the classification tests to the velocity method.  The classification tests are performed by considering a K-S test, a sample set of n=4 simulations and the focal parameters  A) $\hat{P}=(0.1,0.01)$, B) $\hat{P}=(0.5,0.01)$, and C) $\hat{P}=(0.9,0.01)$. In each plot at each parameter pair $(p_m,p_p)$,  circle whose color corresponds to the percentage of times the migration measurements of that parameter pair are statistically significantly different to those for the focal parameter $\hat{P}$. We indicate the focal parameter pair with a red circle. The plots illustrate how the classification performance of the method varies as the motility parameter varies. The method performs better when the motility parameter is small.
   \label{fig:classification_velmethod} }
   \end{figure*}

In Figure~\ref{fig:classification_velmethod}, we plot the results of the mean behaviour of the classification tests when considering the K-S test and the three focal parameter combinations: $\hat{P}=(0.1,0.01)$, $(0.5,0.01)$ and $(0.9,0.01)$. On each plot, the focal parameter combination is indicated by a red circle. At each position  $(p_m,p_p)$, we plot a circle whose color corresponds to the percentage of times the migration measurements of that parameter pair are statistically significantly different to those for the focal parameters $\hat{P}$ with respect to the colorbar at the left of the plots. We can observe that for $\hat{P}=(0.1,0.01)$, the classification is perfect: the K-S test indicated that the windowed velocities from simulations of parameter pairs different from the focal parameter, $(p_m,p_p)\neq \hat{P}$, are statistically significantly different to the windowed velocities from simulations of the focal parameter pair (Figure~\ref{fig:classification_velmethod} A) ) 100\% of the time. For $\hat{P}=(0.5,0.01)$, there are four parameter pairs different to the focal one for which the velocities were 80\%, 85\%, 85\% and 95\% times statistically significantly different to those for the focal parameter  (Figure~\ref{fig:classification_velmethod} B) ), and for  $\hat{P}=(0.9,0.01)$, the number of parameter pairs for which the percentage is not 100\% is increased (Figure~\ref{fig:classification_velmethod} C) ). We observe that as the motility rate increases, the classification performance worsens.

\subsubsection{Comparison with standard migration quantification methods}\label{subsec:comparison_quant_methods}

We compare the classification performance of the velocity method with the closure rate and the area methods  \citep{grada2017research,masuzzo2016taking}. As before, the focal parameter combination, $\hat{P}$, takes values in $\{0.1,0.5,0.9\}\times\{0.01,0.05,0.09\}$.  We consider $n=4$ simulations as the sample size and repeat the classification test 20 times.

\begin{figure*}[!htb]
  \centering
   \includegraphics[width=\linewidth]{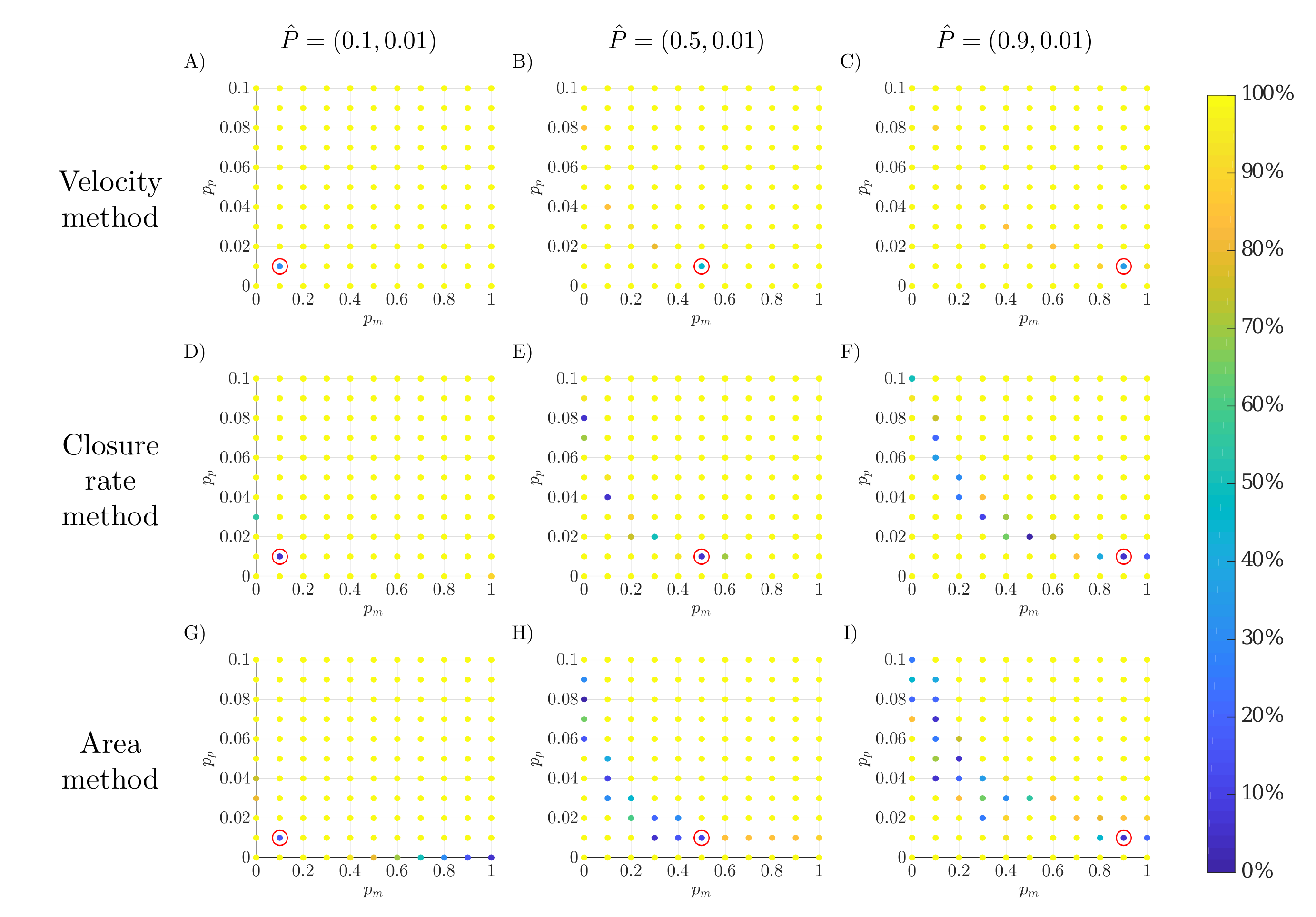}
 \caption{ Series of plots showing how the performance of the three quantification methods changes as the motility rate of the focal parameters varies. In each plot, the color of the circle at each parameter pair $(p_m,p_p)$ indicates the percentage of times the migration measurements associated with the parameter pair are statistically significantly different from those associated with the focal parameters $\hat{P}$. The focal parameters $\hat{P}$ are indicated by a red circle. The results reveal that the velocity method yields a better statistical classification than the other methods. We note also the performance of all three methods declines as the motility rate of the focal parameters $\hat{P}$ increases.   \label{fig:classification_tests1}}
  \end{figure*}

In Figure~\ref{fig:classification_tests1}, we plot the mean behaviour of the classification tests for the three quantification methods by applying the K-S test and the focal parameter combinations $\hat{P}=(0.1,0.01)$, $(0.5,0.01)$ and $(0.9,0.01)$. We observe that for a focal parameter pair, the velocity method yields fewer wrong classifications. We also observe that as the proliferation rate increases, the percentage number of wrong classifications increases for the three methods.

The results of the classification tests for all other focal parameter combinations in  $\{0.1,0.5,0.9\}\times\{0.01,0.05,0.09\}$ are presented in the Supporting Information Section 2. Overall we observed that our method outperforms the closure rate and the area method. For all focal parameter combinations tested, the velocity method yielded greater percentage of correct classifications.  The performance of the area method was the worst while the performance of the closure rate method was intermediate between our method and the area method. The performance of all three methods declines as the motility and the proliferation rate of the focal parameters $\hat{P}$ increase.

\subsection{Application of the quantification methods to in-vitro data}

Having tested the quantification methods on in-silico data, we now use them to analyse experimental data. We first detect the position of the leading edges from the wound healing  images taken during the course of the experiments.  We then quantify the migration rates using the three quantification methods and analyse the statistical classification.

\subsubsection{Image segmentation }

After applying the segmentation algorithm, the front of the cell monolayer is detected for each time-lapse image. The quality of the images was variable, depending on the contrast between the medium and the cell masses, the initial scratch uniformity and the initial curvature of the interfaces. In Figure~\ref{fig:segmentation_example} we present the evolution of a representative scratch from each cell group (S1-S6).

\subsubsection{Quantification method results}

\begin{figure}
\centering
  \includegraphics[width=0.7\linewidth]{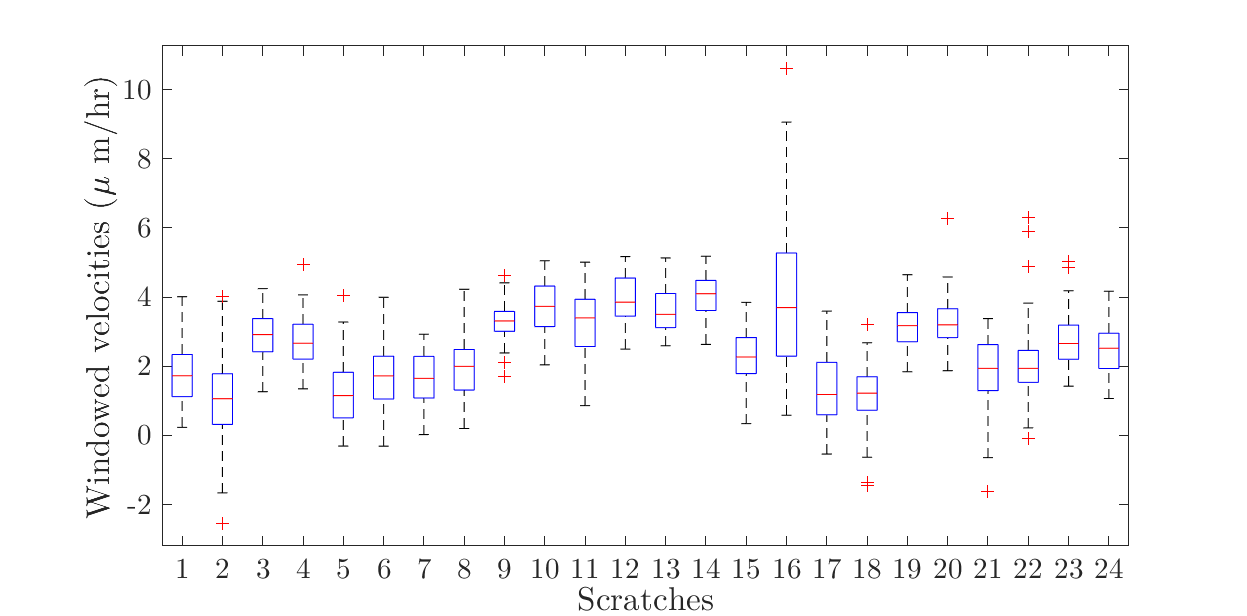}
  \caption{Boxplots of the windowed velocities with respect to the optimal window size 16 for each experimental scratch assay is depicted. \label{fig:windowed_velocities}}
\end{figure}

We quantify the migration velocity of scratch assays for the different cell types using the velocity method. We determine the global optimal window by calculating the objective function for the 24 scratch assays. We vary the window size $w$ from 1 to $500 \mu m$ with a step size of $1 \mu m$ and use equation \eqref{eq:global_fit} to calculate the objective function $F(w)$. The objective function and the three fitness functions that contribute to its calculation are shown in the Supporting Information Section 3. The maximum value is attained for a window size of $16\mu$m. For a fixed window size ($w=16\mu$m), we use a linear approximation to describe the evolution of the fronts and determine the 32 representative windowed velocities for each scratch assay and visualize their boxplots
in Figure~\ref{fig:windowed_velocities}.

We analyse the autocorrelation function of the windowed velocities to determine whether the velocities can be viewed as independent and identically distributed samples in a statistical test. In the Supporting Information Section 4, we include the plots of the autocorrelation functions of the left and right windowed velocities from each scratch assay. We observe that, except for the autocorrelation coefficient at lag 1, all other autocorrelation coefficients are within the 95\% confidence limit for a random independent and identically distributed sequence. Therefore we consider the windowed velocities to be independent and identically distributed samples.

\subsubsection{Statistical classification via the local quantification method}\label{subsec:classification_experimental_data}

After grouping the velocities of scratch assays from the same cell type, the migration rate of each cell group is represented by 264 velocities. The  boxplots associated with the velocity distributions for the six groups are shown in Figure~\ref{fig:boxplot_groups_all} A). To determine how different the migration rate of cell group S1 from the others, we perform a K-S test to test the null hypothesis that the velocities from the two groups come from the same distribution. The null hypothesis was rejected for groups S2, S3 and S4 with statistical significance level of  $p_{value}\leq 0.0001$. The null hypothesis was rejected for group S6 with statistical significance level of  $p_{value}\leq 0.05$. For group S5, the null hypothesis was not rejected. We performed a t-test between S1 and each of the other groups to determine whether the mean difference is statistically significant. The mean difference between the velocities for cell groups S1 and S2, S3 and S4 is statistically significant at the 0.0001 level.  There was statistical significance in the mean difference with respect to S6 at the 0.05 significance level. The statistical results for the K-S tests and t-tests are reported in Figure~\ref{fig:boxplot_groups_all} A). The exact value of the $p_{value}$ for each test is reported in the Supporting Information 5.

\subsubsection{Statistical comparison to standard migration quantification methods}

\begin{figure*}[!htb]
  \centering
  \includegraphics[width=0.9\linewidth]{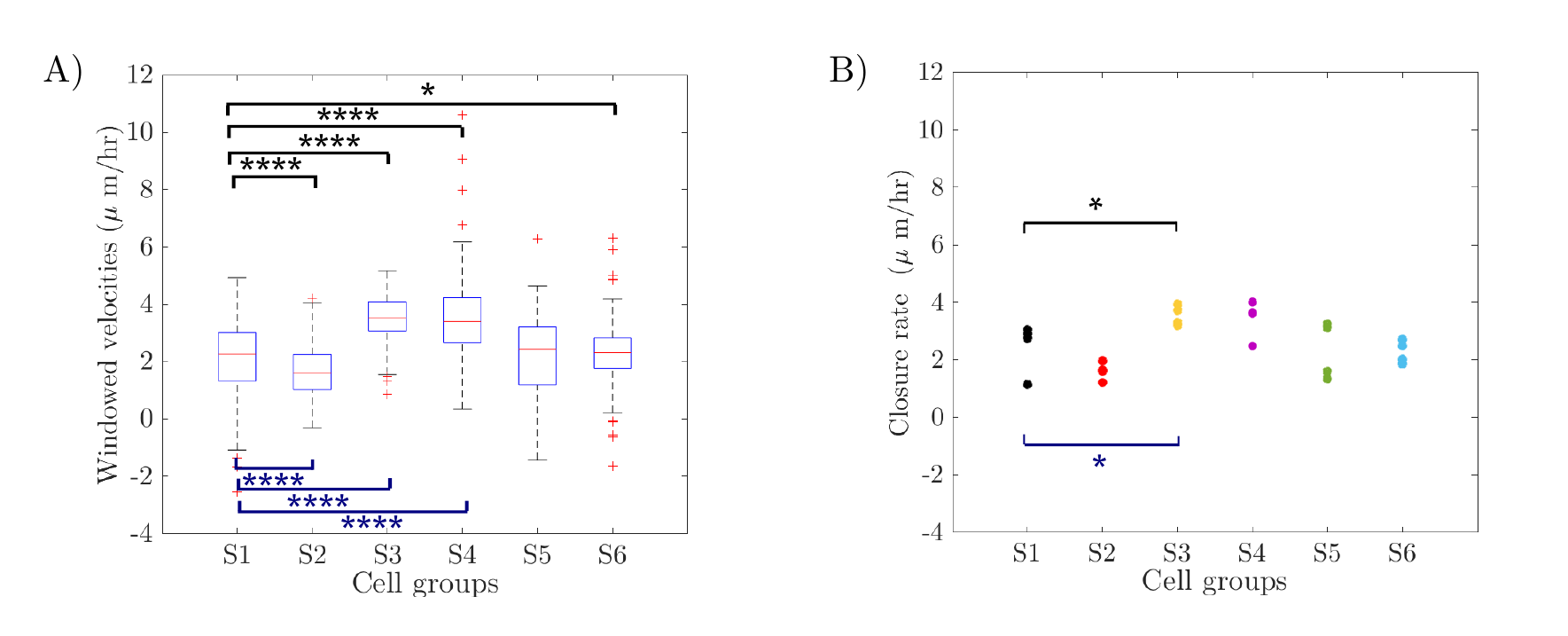}
  \caption{ Statistical analysis of the experimental data using the veloctiy and the closure rate method.  First, the migration measurements are grouped into the six different groups (S1-S6).  The windowed velocties and the closure rates for the cell groups are plotted in  A and B, respectively. Above the data, in black, we have reported the statistical significance results from performing a K-S test with respect to the S1 group. Below the data, we have done the same for the t-tests. Considering the windowed velocities, with respect to the K-S test and t-test, the null hypothesis was rejected testing group S1 against group S2, S3 and S4 at the 0.001 significance level. Performing the statistical tests with the closure rate measurements, the null hypothesis was rejected at the significance level of 0.05 between S1 and S3. The statistical significance level is decoded in the symbols:  *:= $p_{value}\leq 0.05$, **:= $p_{value}\leq 0.01$, ***:= $p_{value}\leq 0.001$ and ****:= $p_{value}\leq 0.0001$ }\label{fig:boxplot_groups_all}
\end{figure*}

We now compare the statistical results of our quantification method against those for the area and closure rate methods. In Figure~\ref{fig:boxplot_groups_all} B) we plot the closure rates of each group and report the results from performing the K-S test and t-test between S1 and the other groups. S3 was the only group for which the null hypothesis of the K-S test and the t-test was rejected at the 0.05 significance level. When we performed the statistical tests for the percentage area measurements, no significant difference was found. In Section 5 of the Supporting Information we include the results of the K-S and t-tests for the percentage wound area measurements.

\section{Discussion}

In this work we have introduced a new migration quantification method for scratch assays that characterizes the x-component of the front velocity of cell monolayers. The method involves three steps: (1) determination of an optimal window $w^*$ with which to approximate the cell front by a function which is piecewise constant in segments of length $w^*$; (2) approximation of the interface with respect to the window size $w^*$ at each time point; and  (3) linear approximation of the evolution of the interface in each of these windows. In this way we characterize cell migration in the scratch assay by the slopes of a series of linear approximations to the interface evolution in these windows. The optimal window is chosen to be the one that best fits a constant velocity profile and for which the left and right velocities belonging to the same distribution.

By developing an agent-based model that mimics the scratch assay, we tested the ability of our quantification method to distinguish between cell lines with known cell motility and proliferation rates. As the motility and proliferation rates increased, the mean and variance of the windowed velocities increased.

By comparing our quantification method with two existing methods, we observed that our method outperforms both in that it yielded a greater percentage of correct classifications than the other methods across a range of parameter values.  Despite being widely used, the performance of the area method was the worst while the performance of the closure rate method was intermediate between our method and the area method. If the leading edges of the initial scratch were perfectly aligned and equidistant from each other, then applying the closure rate method is equivalent to approximating the fronts by their mean positions and the closure rate represents the slope of the linear approximation to the evolution of the interface. The poor performance of the closure rate method is related to the irregularity of the data.

After showing that our quantification method performed better regarding the statistical classification using \textit{in-silico} data, we then used it to analyse our experimental data set. We calculated an optimal window of 16 $\mu m $ and then determined the corresponding windowed horizontal velocities. By performing two sample Kolmogorov-Smirnov and unpaired two-sample t-tests, we identified a statistically significant difference between the S1 group and groups S2-S4. The K-S test also indicated statistically significant differences with respect to group S6. We used these two tests since we wanted to detect differences at the distribution level (through the K-S test) and at the mean level (through the t-test). The closure rate method only detected statistically significant differences between S1 and S3. The closure rate data is of poor quality: more samples are needed to analyse the migration rate with this method. The area method was unable to detect any differences in the data. Even when we tried multiple time points, there was no significant difference.  We observed that the S1 cell group also exhibited the highest levels of expression of target genes associated with malignancy and poor prognosis, when analysed by qRT-PCR techniques (data not shown) in agreement with the detected significant differences in migration.

In conclusion, we have introduced a new quantification framework that outperforms existing quantification methods. Using in silico data, we studied the performance of the new method for a system involving cell proliferation and motility. We compared its performance against two other quantification methods and showed that our method outperforms both of them.  We then applied our quantification method to in-vitro data and found that specific mutations present on this transcription factor have an impact on the migratory capacity, whereas the other methods were unable to detect these important differences. There are several ways our study can be extended. The cell monolayer front evolution can be fitted to a Richards function, a non-symmetrical sigmoid function, which accounts for an initial phase during which the cells react to the presence of the wound, as in \cite{topman2012standardized}. The statistical performance of the quantification method could be tested on publicly available wound healing experiment data sets such as those in \cite{zaritsky2015live}, which provide data from sets of assays and replicates under different experimental conditions. While the Particle Image Velocimetry (PIV) and the Cell Image Velocimetry (CIV) \citep{adrian2011particle,milde2012cell} provides a more in depth description of the velocity field across the cell monolayer, it is more computationally intensive to implement and images need to be recorded at a faster rate to obtain a good approximation to the velocity field. Despite our method not providing the evolution of the velocity field across the full monolayer, its classification performance is superior to common quantification methods. It is important to remember that when performing statistical tests such as the K-S test and t-test between samples of cell populations under different experimental conditions and testing for difference between treatment groups, a small p-value simply flags the data as being unusual under the null hypothesis and assuming all the assumptions of the model were correct \citep{greenland2016statistical}.

\section*{Funding}

This work was supported by the Heidelberg Graduate School of Mathematical and Computational Methods for the Sciences [DFG grant GSC 220 in the German Universities Excellence Initiative to AVPB]; Mathematics for Industry Network [COST Action TD1409 short-term scientific missions grant to AVPB]; Consejo Nacional de Ciencia y Tecnolog\'ia [411678 to JA]; Ministerio de Ciencia e Innovaci\'on [TIN2015-66951-C2-1-R and SGR 1742 to SB, SAF2017-89989-R  to AM]; Red de Investigaci\'on Renal REDinREN [12/0021/0013 to AM]; the CERCA Programme of the Generalitat de Catalunya to TA; Ministerio de Econom\'ia y Competitividad [MTM2015-71509-C2-1-R and MDM-2014-0445 to TA] and  Ag\`encia de Gesti\'o de Ajuts Universitaris i Recerca [2014SGR1307 to TA].
\vspace*{-12pt}

\section*{Acknowledgements}

AVPB would like to thank Isabel Serra Mochales, Matthew Simpson and Andreas Spitz for helpful discussions. We would like to thank Guillem Perez for being the initial promoter of this collaboration.  


\vspace*{-12pt}

\begin{thebibliography}{}

\bibitem[Adrian and Westerweel(2011)Adrian and Westerweel]{adrian2011particle}
Adrian, R.~J. and Westerweel, J. (2011).
\newblock {\em Particle image velocimetry\/}.
\newblock Number~30. Cambridge University Press.

\bibitem[B{\"u}th {\em et~al.}(2007)B{\"u}th, Buttigieg, Ostafe, Rehders,
  Dannenmann, Schaschke, Stark, Boukamp, and Brix]{buth2007cathepsin}
B{\"u}th, H., Buttigieg, P.~L., Ostafe, R., Rehders, M., Dannenmann, S.~R.,
  Schaschke, N., Stark, H.-J., Boukamp, P., and Brix, K. (2007).
\newblock Cathepsin b is essential for regeneration of scratch-wounded normal
  human epidermal keratinocytes.
\newblock {\em European journal of cell biology\/}, {\bf 86}(11-12), 747--761.

\bibitem[Canny(1987)Canny]{canny1987computational}
Canny, J. (1987).
\newblock A computational approach to edge detection.
\newblock In {\em Readings in Computer Vision\/}, pages 184--203. Elsevier.

\bibitem[Ferreira and Rasband(2012)Ferreira and Rasband]{ferreira2012imagej}
Ferreira, T. and Rasband, W. (2012).
\newblock Imagej user guide.
\newblock {\em ImageJ/Fiji\/}, {\bf 1}.

\bibitem[Friedl and Gilmour(2009)Friedl and Gilmour]{friedl2009collective}
Friedl, P. and Gilmour, D. (2009).
\newblock Collective cell migration in morphogenesis, regeneration and cancer.
\newblock {\em Nature reviews Molecular cell biology\/}, {\bf 10}(7), 445.

\bibitem[Gebaeck(2009)Gebaeck]{gebaeck2009tscratch}
Gebaeck (2009).
\newblock Tscratch: a novel and simple software tool for automated analysis of
  monolayer wound healing assays (vol 46, pg 265, 2009).
\newblock {\em BioTechniques\/}, {\bf 46}(6), 383--383.

\bibitem[Gorshkova {\em et~al.}(2008)Gorshkova, He, Berdyshev, Usatuyk, Burns,
  Kalari, Zhao, Pendyala, Garcia, Pyne, {\em et~al.}]{gorshkova2008protein}
Gorshkova, I., He, D., Berdyshev, E., Usatuyk, P., Burns, M., Kalari, S., Zhao,
  Y., Pendyala, S., Garcia, J.~G., Pyne, N.~J., {\em et~al.} (2008).
\newblock Protein kinase c-ϵ regulates sphingosine 1-phosphate-mediated
  migration of human lung endothelial cells through activation of phospholipase
  d2, protein kinase c-$\zeta$, and rac1.
\newblock {\em Journal of Biological Chemistry\/}, {\bf 283}(17), 11794--11806.

\bibitem[Grada {\em et~al.}(2017)Grada, Otero-Vinas, Prieto-Castrillo, Obagi,
  and Falanga]{grada2017research}
Grada, A., Otero-Vinas, M., Prieto-Castrillo, F., Obagi, Z., and Falanga, V.
  (2017).
\newblock Research techniques made simple: Analysis of collective cell
  migration using the wound healing assay.
\newblock {\em Journal of Investigative Dermatology\/}, {\bf 137}(2), e11--e16.

\bibitem[Greenland {\em et~al.}(2016)Greenland, Senn, Rothman, Carlin, Poole,
  Goodman, and Altman]{greenland2016statistical}
Greenland, S., Senn, S.~J., Rothman, K.~J., Carlin, J.~B., Poole, C., Goodman,
  S.~N., and Altman, D.~G. (2016).
\newblock Statistical tests, p values, confidence intervals, and power: a guide
  to misinterpretations.
\newblock {\em European journal of epidemiology\/}, {\bf 31}(4), 337--350.

\bibitem[Heiberger and Holland(2015)Heiberger and
  Holland]{heiberger2015statistical}
Heiberger, R.~M. and Holland, B. (2015).
\newblock {\em Statistical analysis and data display: an intermediate course
  with examples in R\/}.
\newblock Springer.

\bibitem[Hulkower and Herber(2011)Hulkower and Herber]{hulkower2011cell}
Hulkower, K.~I. and Herber, R.~L. (2011).
\newblock Cell migration and invasion assays as tools for drug discovery.
\newblock {\em Pharmaceutics\/}, {\bf 3}(1), 107--124.

\bibitem[Jin {\em et~al.}(2016)Jin, Shah, Penington, McCue, Chopin, and
  Simpson]{jin2016reproducibility}
Jin, W., Shah, E.~T., Penington, C.~J., McCue, S.~W., Chopin, L.~K., and
  Simpson, M.~J. (2016).
\newblock Reproducibility of scratch assays is affected by the initial degree
  of confluence: experiments, modelling and model selection.
\newblock {\em Journal of theoretical biology\/}, {\bf 390}, 136--145.

\bibitem[Jin {\em et~al.}(2017)Jin, Lo, Chou, McCue, and Simpson]{jin2017role}
Jin, W., Lo, K.-Y., Chou, S.-E., McCue, S.~W., and Simpson, M.~J. (2017).
\newblock The role of initial geometry in experimental models of wound closing.
\newblock {\em arXiv preprint arXiv:1711.07162\/}.

\bibitem[Johnston {\em et~al.}(2014)Johnston, Simpson, and
  McElwain]{johnston2014much}
Johnston, S.~T., Simpson, M.~J., and McElwain, D.~S. (2014).
\newblock How much information can be obtained from tracking the position of
  the leading edge in a scratch assay?
\newblock {\em Journal of the Royal Society Interface\/}, {\bf 11}(97),
  20140325.

\bibitem[Johnston {\em et~al.}(2016)Johnston, Ross, Binder, McElwain, Haridas,
  and Simpson]{johnston2016quantifying}
Johnston, S.~T., Ross, J.~V., Binder, B.~J., McElwain, D.~S., Haridas, P., and
  Simpson, M.~J. (2016).
\newblock Quantifying the effect of experimental design choices for in vitro
  scratch assays.
\newblock {\em Journal of theoretical biology\/}, {\bf 400}, 19--31.

\bibitem[Jonkman {\em et~al.}(2014)Jonkman, Cathcart, Xu, Bartolini, Amon,
  Stevens, and Colarusso]{jonkman2014introduction}
Jonkman, J.~E., Cathcart, J.~A., Xu, F., Bartolini, M.~E., Amon, J.~E.,
  Stevens, K.~M., and Colarusso, P. (2014).
\newblock An introduction to the wound healing assay using live-cell
  microscopy.
\newblock {\em Cell adhesion \& migration\/}, {\bf 8}(5), 440--451.

\bibitem[Khain {\em et~al.}(2011)Khain, Katakowski, Hopkins, Szalad, Zheng,
  Jiang, and Chopp]{khain2011collective}
Khain, E., Katakowski, M., Hopkins, S., Szalad, A., Zheng, X., Jiang, F., and
  Chopp, M. (2011).
\newblock Collective behavior of brain tumor cells: the role of hypoxia.
\newblock {\em Physical Review E\/}, {\bf 83}(3), 031920.

\bibitem[Kramer {\em et~al.}(2013)Kramer, Walzl, Unger, Rosner, Krupitza,
  Hengstschl{\"a}ger, and Dolznig]{kramer2013vitro}
Kramer, N., Walzl, A., Unger, C., Rosner, M., Krupitza, G., Hengstschl{\"a}ger,
  M., and Dolznig, H. (2013).
\newblock In vitro cell migration and invasion assays.
\newblock {\em Mutation Research/Reviews in Mutation Research\/}, {\bf 752}(1),
  10--24.

\bibitem[Li {\em et~al.}(2013)Li, He, Zhao, and Jiang]{li2013collective}
Li, L., He, Y., Zhao, M., and Jiang, J. (2013).
\newblock Collective cell migration: Implications for wound healing and cancer
  invasion.
\newblock {\em Burns \& trauma\/}, {\bf 1}(1), 21.

\bibitem[Liang {\em et~al.}(2007)Liang, Park, and Guan]{liang2007vitro}
Liang, C.-C., Park, A.~Y., and Guan, J.-L. (2007).
\newblock In vitro scratch assay: a convenient and inexpensive method for
  analysis of cell migration in vitro.
\newblock {\em Nature protocols\/}, {\bf 2}(2), 329.

\bibitem[Maini {\em et~al.}(2004)Maini, McElwain, and
  Leavesley]{maini2004traveling}
Maini, P.~K., McElwain, D.~S., and Leavesley, D.~I. (2004).
\newblock Traveling wave model to interpret a wound-healing cell migration
  assay for human peritoneal mesothelial cells.
\newblock {\em Tissue engineering\/}, {\bf 10}(3-4), 475--482.

\bibitem[Masuzzo {\em et~al.}(2016)Masuzzo, Van~Troys, Ampe, and
  Martens]{masuzzo2016taking}
Masuzzo, P., Van~Troys, M., Ampe, C., and Martens, L. (2016).
\newblock Taking aim at moving targets in computational cell migration.
\newblock {\em Trends in cell biology\/}, {\bf 26}(2), 88--110.

\bibitem[Milde {\em et~al.}(2012)Milde, Franco, Ferrari, Kurtcuoglu,
  Poulikakos, and Koumoutsakos]{milde2012cell}
Milde, F., Franco, D., Ferrari, A., Kurtcuoglu, V., Poulikakos, D., and
  Koumoutsakos, P. (2012).
\newblock Cell image velocimetry (civ): boosting the automated quantification
  of cell migration in wound healing assays.
\newblock {\em Integrative Biology\/}, {\bf 4}(11), 1437--1447.

\bibitem[{\"O}ner and Deveci~Kocako{\c{c}}(2017){\"O}ner and
  Deveci~Kocako{\c{c}}]{oner2017jmasm}
{\"O}ner, M. and Deveci~Kocako{\c{c}}, {\.I}. (2017).
\newblock Jmasm 49: A compilation of some popular goodness of fit tests for
  normal distribution: Their algorithms and matlab codes (matlab).
\newblock {\em Journal of Modern Applied Statistical Methods\/}, {\bf 16}(2),
  30.

\bibitem[Ranzato {\em et~al.}(2011)Ranzato, Martinotti, and
  Burlando]{ranzato2011wound}
Ranzato, E., Martinotti, S., and Burlando, B. (2011).
\newblock Wound healing properties of jojoba liquid wax: an in vitro study.
\newblock {\em Journal of ethnopharmacology\/}, {\bf 134}(2), 443--449.

\bibitem[Roussos {\em et~al.}(2011)Roussos, Condeelis, and
  Patsialou]{roussos2011chemotaxis}
Roussos, E.~T., Condeelis, J.~S., and Patsialou, A. (2011).
\newblock Chemotaxis in cancer.
\newblock {\em Nature Reviews Cancer\/}, {\bf 11}(8), 573.

\bibitem[Ruiz-Ca{\~n}ada {\em et~al.}(2017)Ruiz-Ca{\~n}ada,
  Bernab{\'e}-Garc{\'\i}a, Liarte, Insausti, Angosto, Moraleda, Castellanos,
  and Nicol{\'a}s]{ruiz2017amniotic}
Ruiz-Ca{\~n}ada, C., Bernab{\'e}-Garc{\'\i}a, {\'A}., Liarte, S., Insausti,
  C.~L., Angosto, D., Moraleda, J.~M., Castellanos, G., and Nicol{\'a}s, F.~J.
  (2017).
\newblock Amniotic membrane stimulates cell migration by modulating
  transforming growth factor-$\beta$ signaling.
\newblock {\em Journal of tissue engineering and regenerative medicine\/}.

\bibitem[Shrivakshan {\em et~al.}(2012)Shrivakshan, Chandrasekar, {\em
  et~al.}]{shrivakshan2012comparison}
Shrivakshan, G., Chandrasekar, C., {\em et~al.} (2012).
\newblock A comparison of various edge detection techniques used in image
  processing.
\newblock {\em IJCSI International Journal of Computer Science Issues\/}, {\bf
  9}(5), 272--276.

\bibitem[Simpson {\em et~al.}(2008)Simpson, Selfors, Bui, Reynolds, Leake,
  Khvorova, and Brugge]{simpson2008identification}
Simpson, K.~J., Selfors, L.~M., Bui, J., Reynolds, A., Leake, D., Khvorova, A.,
  and Brugge, J.~S. (2008).
\newblock Identification of genes that regulate epithelial cell migration using
  an sirna screening approach.
\newblock {\em Nature cell biology\/}, {\bf 10}(9), 1027.

\bibitem[Simpson {\em et~al.}(2010)Simpson, Landman, and
  Hughes]{simpson2010cell}
Simpson, M.~J., Landman, K.~A., and Hughes, B.~D. (2010).
\newblock Cell invasion with proliferation mechanisms motivated by time-lapse
  data.
\newblock {\em Physica A: Statistical Mechanics and its Applications\/}, {\bf
  389}(18), 3779--3790.

\bibitem[Simpson {\em et~al.}(2013)Simpson, Treloar, Binder, Haridas, Manton,
  Leavesley, McElwain, and Baker]{simpson2013quantifying}
Simpson, M.~J., Treloar, K.~K., Binder, B.~J., Haridas, P., Manton, K.~J.,
  Leavesley, D.~I., McElwain, D.~S., and Baker, R.~E. (2013).
\newblock Quantifying the roles of cell motility and cell proliferation in a
  circular barrier assay.
\newblock {\em Journal of the Royal Society Interface\/}, {\bf 10}(82),
  20130007.

\bibitem[Tisue and Wilensky(2004)Tisue and Wilensky]{tisue2004netlogo}
Tisue, S. and Wilensky, U. (2004).
\newblock Netlogo: A simple environment for modeling complexity.
\newblock In {\em International conference on complex systems\/}, volume~21,
  pages 16--21. Boston, MA.

\bibitem[Topman {\em et~al.}(2012)Topman, Sharabani-Yosef, and
  Gefen]{topman2012standardized}
Topman, G., Sharabani-Yosef, O., and Gefen, A. (2012).
\newblock A standardized objective method for continuously measuring the
  kinematics of cultures covering a mechanically damaged site.
\newblock {\em Medical Engineering and Physics\/}, {\bf 34}(2), 225--232.

\bibitem[Treloar and Simpson(2013)Treloar and Simpson]{treloar2013sensitivity}
Treloar, K.~K. and Simpson, M.~J. (2013).
\newblock Sensitivity of edge detection methods for quantifying cell migration
  assays.
\newblock {\em PloS one\/}, {\bf 8}(6), e67389.

\bibitem[Vargas {\em et~al.}(2015)Vargas, Angeli, Pastrello, McQuaid, Li,
  Jurisicova, and Jurisica]{vargas2015robust}
Vargas, A., Angeli, M., Pastrello, C., McQuaid, R., Li, H., Jurisicova, A., and
  Jurisica, I. (2015).
\newblock Robust quantitative scratch assay.
\newblock {\em Bioinformatics\/}, {\bf 32}(9), 1439--1440.

\bibitem[Vezhnevets and Konouchine(2005)Vezhnevets and
  Konouchine]{vezhnevets2005growcut}
Vezhnevets, V. and Konouchine, V. (2005).
\newblock Growcut: Interactive multi-label nd image segmentation by cellular
  automata.
\newblock In {\em proc. of Graphicon\/}, volume~1, pages 150--156. Citeseer.

\bibitem[Walter {\em et~al.}(2010)Walter, Wright, Fuller, MacNeil, and
  Johnson]{walter2010mesenchymal}
Walter, M., Wright, K.~T., Fuller, H., MacNeil, S., and Johnson, W. E.~B.
  (2010).
\newblock Mesenchymal stem cell-conditioned medium accelerates skin wound
  healing: an in vitro study of fibroblast and keratinocyte scratch assays.
\newblock {\em Experimental cell research\/}, {\bf 316}(7), 1271--1281.

\bibitem[Zaritsky {\em et~al.}(2011)Zaritsky, Natan, Horev, Hecht, Wolf,
  Ben-Jacob, and Tsarfaty]{zaritsky2011cell}
Zaritsky, A., Natan, S., Horev, J., Hecht, I., Wolf, L., Ben-Jacob, E., and
  Tsarfaty, I. (2011).
\newblock Cell motility dynamics: a novel segmentation algorithm to quantify
  multi-cellular bright field microscopy images.
\newblock {\em PloS one\/}, {\bf 6}(11), e27593.

\bibitem[Zaritsky {\em et~al.}(2013)Zaritsky, Manor, Wolf, Ben-Jacob, and
  Tsarfaty]{zaritsky2013benchmark}
Zaritsky, A., Manor, N., Wolf, L., Ben-Jacob, E., and Tsarfaty, I. (2013).
\newblock Benchmark for multi-cellular segmentation of bright field microscopy
  images.
\newblock {\em BMC bioinformatics\/}, {\bf 14}(1), 319.

\bibitem[Zaritsky {\em et~al.}(2015)Zaritsky, Natan, Kaplan, Ben-Jacob, and
  Tsarfaty]{zaritsky2015live}
Zaritsky, A., Natan, S., Kaplan, D., Ben-Jacob, E., and Tsarfaty, I. (2015).
\newblock Live time-lapse dataset of in vitro wound healing experiments.
\newblock {\em GigaScience\/}, {\bf 4}(1), 8.

\end{thebibliography}
\end{document}